\documentclass[letterpaper,11pt]{article}
\usepackage[margin=25mm]{geometry}
\usepackage{amsmath}
\usepackage{amsfonts}
\usepackage{amssymb}
\usepackage{graphicx}
\usepackage[
  backend=bibtex, 
  mcite=true,
  style=phys,
  articletitle=false,
  biblabel=brackets,
  giveninits=true,
  abbreviate=true,
  doi=false, url=false, isbn=false, eprint=true,
  block=space
]{biblatex}

\usepackage[autostyle=true]{csquotes}

\title{A multistage framework for studying the 
evolution of jets and high-$p_T$ probes in small 
collision systems}

\author{Abhijit Majumder\footnote{majumder@wayne.edu}\,  for The JETSCAPE Collaboration, \\
Department of Physics and Astronomy, Wayne State University,\\
  666 W. Hancock St., Detroit, Michigan, USA}

\begin{document}
\maketitle

\begin{abstract}
Understanding the modification of jets and high-$p_T$ probes in small systems requires the 
integration of soft and hard physics. 
We present recent developments in extending the 
JETSCAPE framework to build an event generator, which includes correlations between soft 
and hard partons, to study jet observables in small systems. 
The multi-scale physics of the 
collision is separated into different stages. Hard scatterings are first sampled at binary 
collision positions provided by the Glauber geometry. They are then propagated backward in
space-time following an initial-state shower to obtain the initiating partons’ energies and 
momenta before the collision. These energies and momenta are then subtracted from the 
incoming colliding nucleons for soft-particle production, modeled by the 3D-Glauber + 
hydrodynamics + hadronic transport framework. 
This new hybrid approach (X-SCAPE) includes non-trivial correlations between jet and soft particle productions in small systems. We calibrate 
this framework with the final state hadron’s $p_T$-spectra from low to 
high $p_T$ in $p$-$p$, and and then compare with the spectra in $p$-$Pb$ collisions from the LHC. We also
present results for additional observables such as the distributions of event activity as a function of the hardest jet $p_T$ in forward and mid-rapidity for both $p$-$p$ and $p$-$Pb$ collisions.
\end{abstract}

\section{Introduction}

Over the last decade there has been extensive interest in the possibility that a quark-gluon plasma (QGP) could have been produced in small collision systems such as $p$-$p$ and $p$-$A$, especially in events with very high multiplicity at top RHIC and LHC energies. The first indication of this was the observation of a ridge in two particle correlations over long ranges in rapidity in both $ p$-$p$ and $p$-$A$ collisions~\mcite{ridge,*ALICE:2012eyl,*ATLAS:2012cix,*CMS:2010ifv,*CMS:2012qk,*PHENIX:2013ktj,*PHENIX:2014fnc,*STAR:2015kak}. The observation of collective flow in $p$-$Pb$ collisions at the LHC~\mcite{v2,*ATLAS:2013jmi,*ALICE:2014dwt,*CMS:2015yux}, reinforced via multi-particle correlations, along with successful theory comparisons, strongly suggests the formation of expanding droplets of QGP in these collisions.  

The \emph{standard model} of heavy-ion collisions~\mcite{reviews,*Muller:2012zq,*Jacobs:2004qv,*Gyulassy:2007kla,*Arslandok:2023utm} includes  the modification of hard jets as a signature of the QGP. This raises the question of the effect on jets (or hard processes in general) from these small droplets of QGP formed in $p$-$p$ and $p$-$Pb$ collisions. 
A study of the nuclear modification factor $R_{pA}$ for both jets and leading hadrons, in minimum bias events, contains no observable  modification, i.e., $R_{pA} \simeq 1$~\cite{ATLAS:2014cpa, CMS:2016xef}. However, if events with jets are binned in centrality, different measures of centrality yield differing variations of $R_{pA}$ with centrality, with some measures yielding a large suppression $R_{pA} \lesssim 1$ for nominally central events, coupled with an enhancement $R_{pA} \gtrsim 1$ for peripheral events~\cite{ATLAS:2014cpa}. Along with this, one also observes a residual azimuthal anisotropy at high $p_T$, even above $20$~GeV, indicative of jet modification~\mcite{highptv2,*ATLAS:2014qaj,*ATLAS:2019vcm}.

A possible reason for these conflicting results could simply be consequences of small system size and the much tighter constraints of energy momentum conservation in the initial stages of such systems. In the typical heavy-ion simulation~\mcite{jetscapejets,*JETSCAPE:2021ehl,*JETSCAPE:2022jer,*JETSCAPE:2022hcb,*JETSCAPE:2023hqn}, one simulates the bulk medium evolution first~\mcite{jetscapebulk,*JETSCAPE:2020mzn,*JETSCAPE:2020shq}. Given the fluctuating configuration of the initial state, a certain amount of energy and momentum is deposited over a range of space-time, and then evolved using a viscous fluid dynamical simulation. After this, one introduces hard jets whose location depends on the binary collision profile and these then propagate and interact with the pre-calculated fluid medium. 
However, no attempt is made to balance the energy-momentum required for the production of the jet with the energy momentum deposited towards the production of the bulk medium, as the  energy within a jet is a tiny fraction of the energy within the bulk medium. This is no longer the case for small collision systems~\cite{Kordell:2016njg}.

In spite of the flexibility of the JETSCAPE framework~\mcite{jetscape-framework,*JETSCAPE:2017eso,*Putschke:2019yrg}, 
it cannot be used to accurately simulate the collisions of small systems where exact initial state conservation of energy-momentum may be required, i.e., the framework does not provide tools to carry this out easily. This situation is ameliorated by the new X-SCAPE framework, which provides various infrastructure enhancements to aid in the incorporation of exact energy-momentum conservation and other initial state effects required in small systems.
In Sec.~\ref{sec:XSCAPE}, we will describe the salient features of the X-SCAPE project. Preliminary results using the new coupled modules of i-MATTER for the hard sector and 3D MC-Glauber for the soft sector are described in Sec.~\ref{sec:new-modules}. An outlook is presented in Sec.~\ref{sec:summary}.

\section{The X-SCAPE framework}
\label{sec:XSCAPE}

The X-ion collisions with a Statistically and Computationally Advanced Program Envelope (X-SCAPE) is the second project of the JETSCAPE collaboration~\cite{XSCAPE}. The main goals of the new framework are to create a JETSCAPE like framework, which is fully backwards compatible, can be extended to encompass the physics of small collision systems, low energy heavy-ion collisions relevant to the beam energy scan, and make the first extension to electron-ion collisions.

\begin{figure}[htb]
\begin{center}
\includegraphics[width=0.8\textwidth]{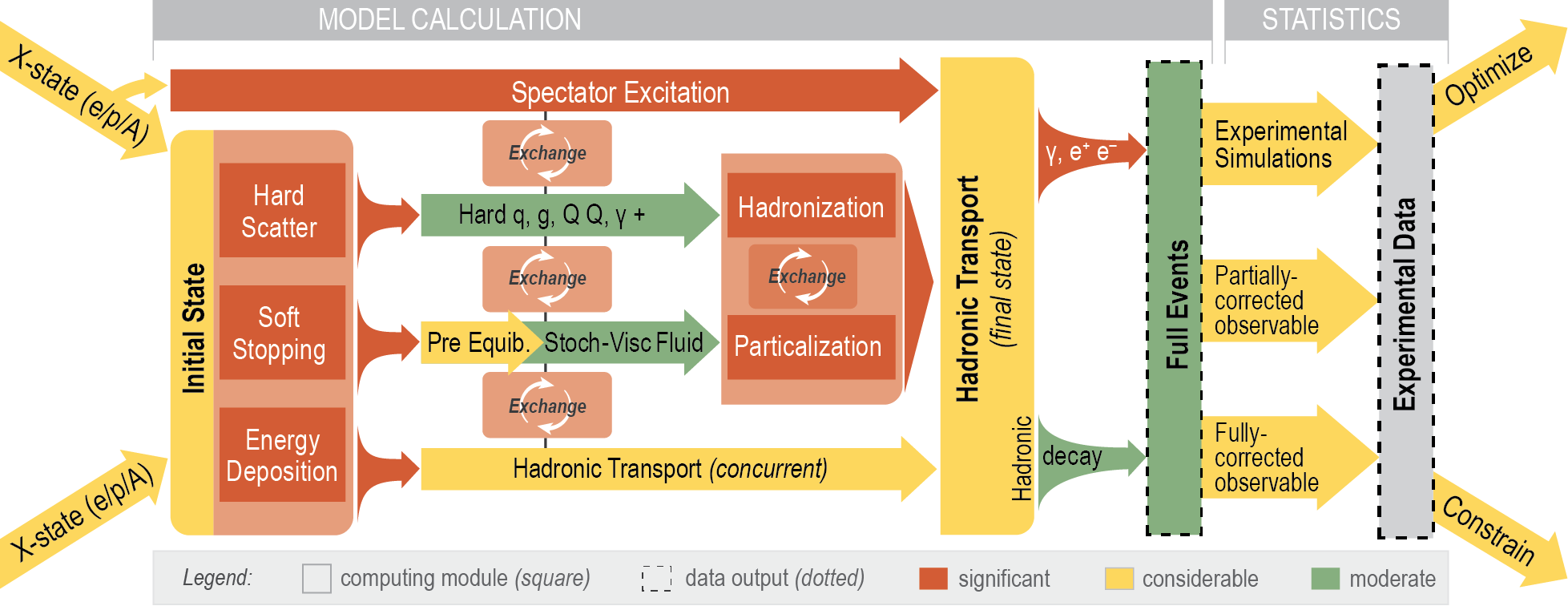} 
\end{center}
\vspace{-0.5cm}
\caption{The flow chart of the X-SCAPE framework}
\label{fig:XSCAPE}
\end{figure}

In Fig.~\ref{fig:XSCAPE}, we present the flow chart of the X-SCAPE framework. Two important differences with the older JETSCAPE framework should immediately become obvious. The input systems are not limited to be $p$-$p$ or $A$-$A$ and can accommodate any system. Also, there is an intentional correlation between the hard scattering, soft stopping and energy deposition modules, connected to the initial state (leftmost red rectangle in Fig.~\ref{fig:XSCAPE}). These will require new modules for both the soft energy deposition and the hard scattering, which can correlate the energy-momentum being deposited in these systems, described in the next section. 

The other new aspects of the new framework are the concurrent running of the hadronic cascade (bottom yellow arrow in Fig.~\ref{fig:XSCAPE}) in tandem with the fluid dynamical simulation (middle green arrow). This aspect is required for the future simulation of lower energy heavy-ion collisions. The other new aspect is the study of the spectator (top red arrow), which is a required aspect of future simulations of electron ion collisions. In the remainder of these proceedings, we will focus on the simulations of small systems using this new framework.

\section{Results with the new modules: i-MATTER and 3D MC-Glauber}
\label{sec:new-modules}

In order to simulate the physics of jet and bulk correlations in $p$-$p$ and $p$-$A$ collisions, we introduce two new modules: 3D MC-Glauber, which generates the bulk or soft energy and Baryon number deposition, setting up the initial state for the subsequent fluid simulation, and i-MATTER which generates the configuration of hard partons prior to the hard scattering.

\begin{figure}[htb]
    \centering
    \includegraphics[width=0.45\textwidth]{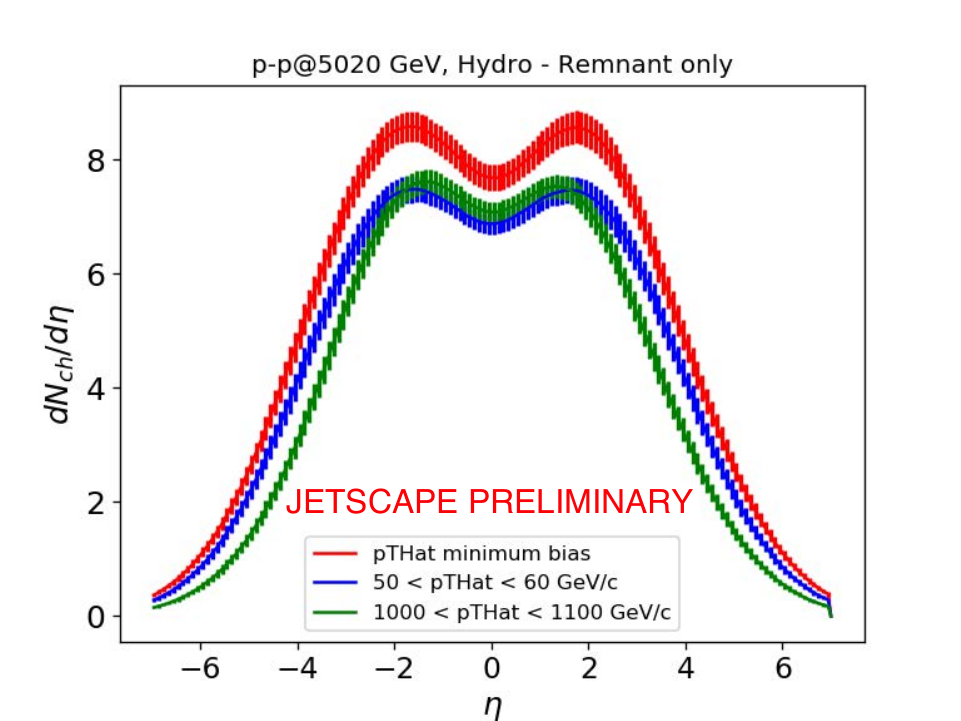}
    \vspace{-0.3cm}
    \caption{Particle production from the bulk simulation with and without a hard scattering in the event.}
    \label{fig:hydro-output}
\end{figure}

Using the X-SCAPE framework, we have introduced a new methodology for simulating small systems in $p$-$Pb$ and $p$-$p$ collisions: The soft sector of all colliding systems, ranging from low multiplicity $p$-$p$ to the highest multiplicity $p$-$A$ (and obviously extending to all $A$-$A$) collisions will be simulated by a viscous fluid dynamical simulation. While fluid dynamics represents actual physics in dense systems such as $A$-$A$ or high multiplicity $p$-$A$, its custodial energy-momentum conservation conditions should provide a reliable model for a majority of soft observables in $p$-$p$. 

\begin{figure}[htb]
    \centering
    \includegraphics[width=0.4\textwidth]{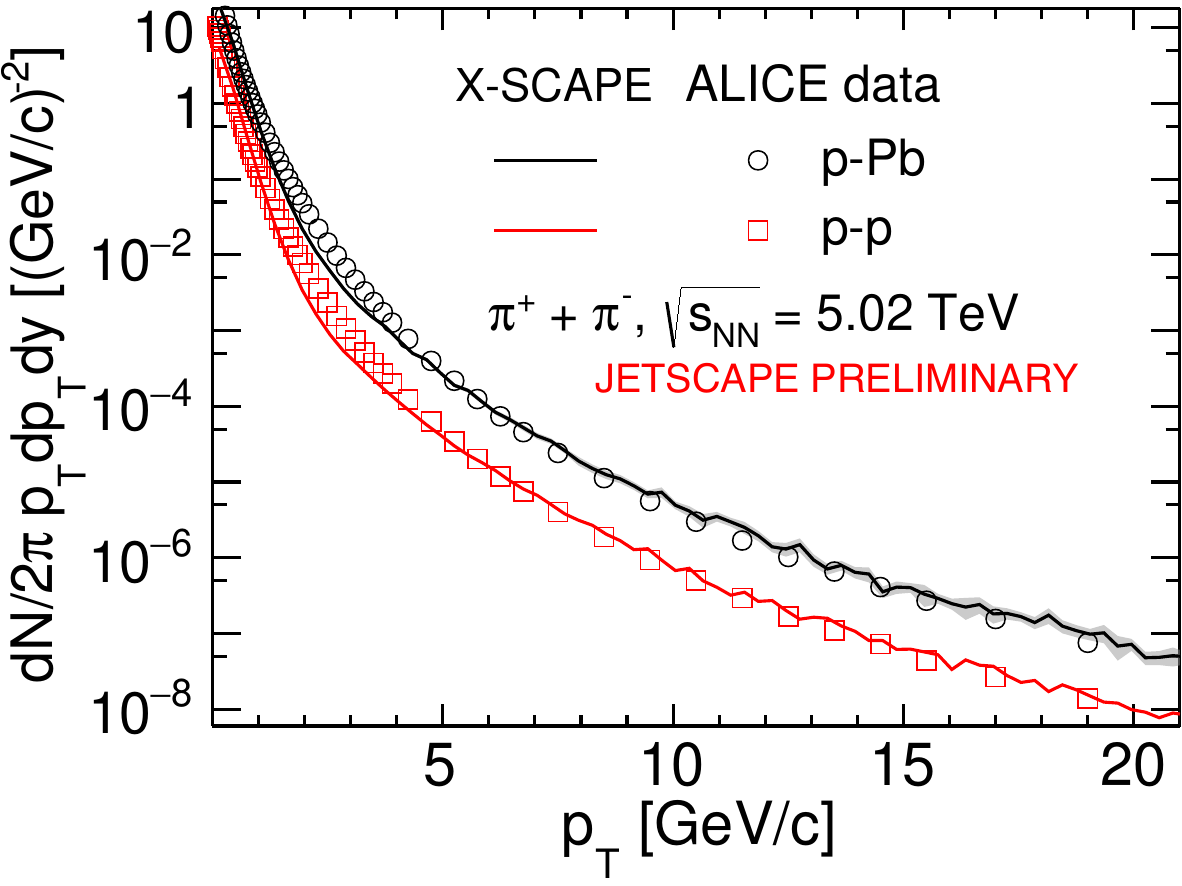}
    \hspace{1cm}
    \includegraphics[width=0.4\textwidth]{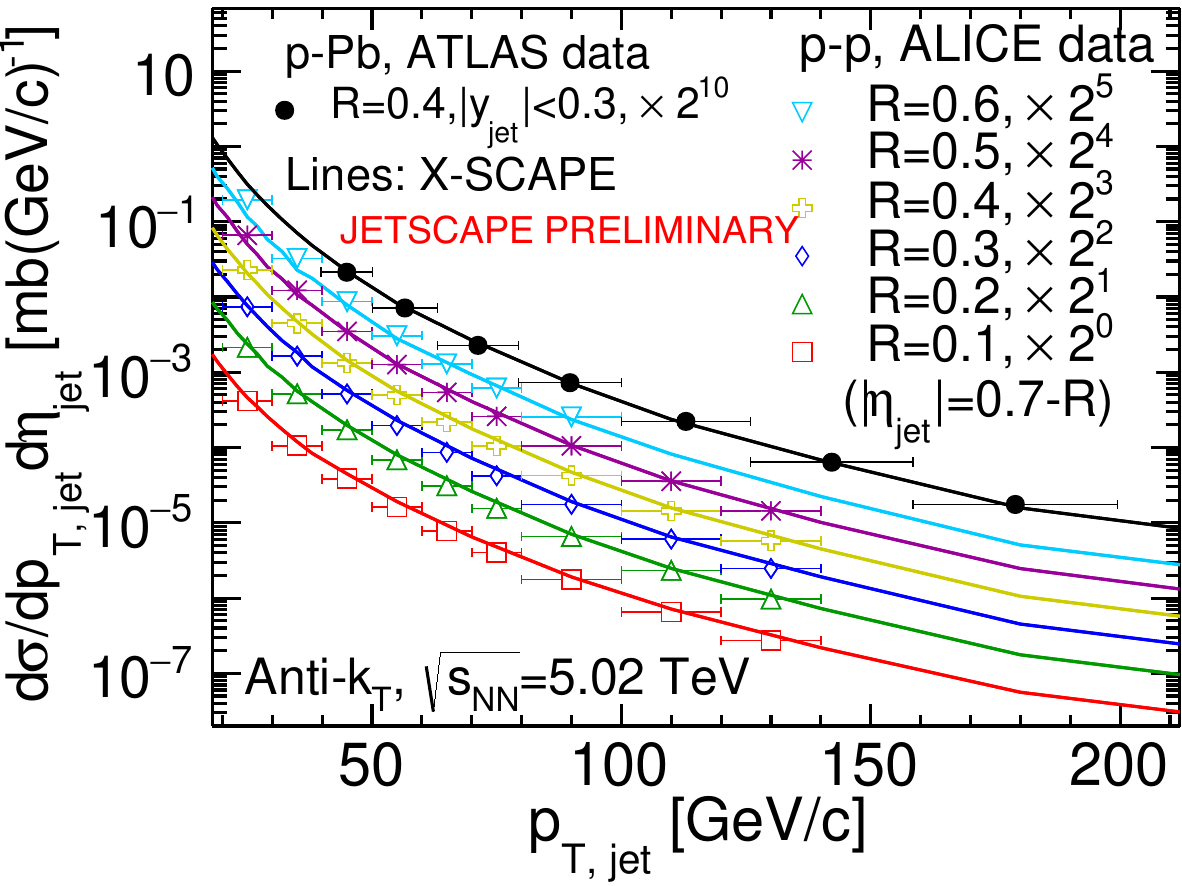}
    \vspace{-0.3cm}
    \caption{Left: Charged pion spectra from $p$-$p$ and $p$-$A$ collisions at $\sqrt{s}=5.02$~TeV as measured by the ALICE experiment compared with preliminary calculations using i-MATTER, PYTHIA and MATTER (final state), along with 3D MC-Glauber and MUSIC within the X-SCAPE framework. Right: Jet spectra measured in the same experiment and compared to the output from the hard sector from X-SCAPE.}
    \label{fig:hadron-jet-spectra}
\end{figure}

The above paradigm successfully reproduces a variety of observables in $p$-$p$ and $p$-$A$ collisions, when combined with a parametrized initial state such as 3D MC-Glauber, as used in this work. The 3D MC-Glauber model~\mcite{MCG,*Shen:2017bsr,*Shen:2022oyg} simulates each incoming proton in terms of several (typically 4) hot-spots. Interactions between these opposing hot spots, calculated using sub-nucleon cross sections generates flux tube-like strings whose spatial extent simulates the deposited energy and baryon number. These energy and baryon number distributions become inputs to the subsequent fluid dynamical simulation simulated using the MUSIC generator. Hadronization of the soft sector is carried out using an event-by-event Cooper-Frye routine called iSS. 

\begin{figure}[htb]
    \centering
    \includegraphics[width=0.4\textwidth]{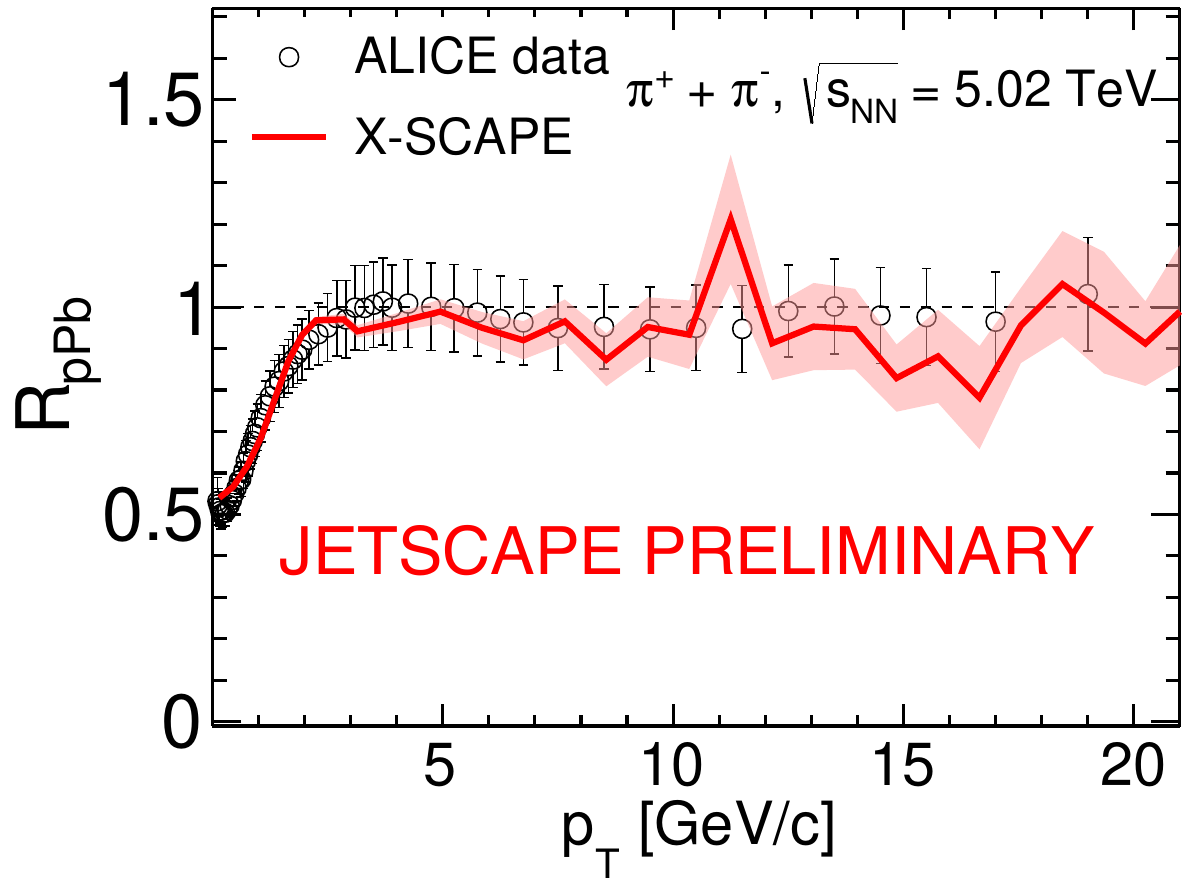}
    \hspace{1cm}
    \includegraphics[width=0.4\textwidth]{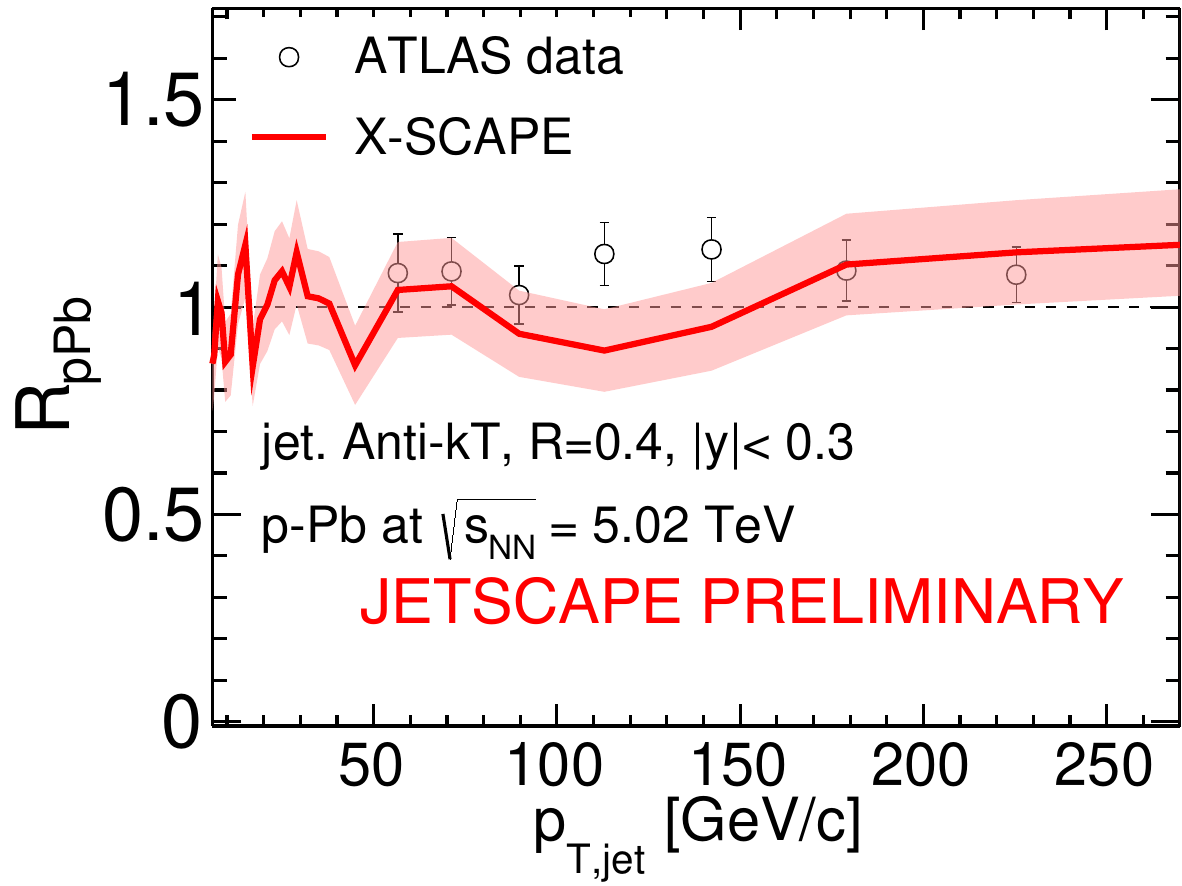}
    \vspace{-0.3cm}
    \caption{Left: Nuclear modification factor for charged hadrons for minimum bias $p$-$Pb$ collisions at $\sqrt{s}=5.02$~TeV compared to preliminary results from the X-SCAPE framework. Right: Nuclear modification factor for jets in the same experiment compared to resutls from X-SCAPE. Spectra are presented in Fig.~\ref{fig:hadron-jet-spectra}.}
    \label{fig:hadron-jet-raa}
\end{figure}

Hard processes are simulated in 4 steps: Hard scatterings are sampled using PYTHIA at locations obtained from sampling the binary hot-spot collision profile from 3D MC-Glauber. Then the initial state radiation is done with i-MATTER, which maintains exact energy-momentum conservation in the backwards shower, along with locations of each splits and the correct color tags for all partons. The initial partons at a large negative time are matched with a remnant anti-parton with the requisite anti-color so that the entire hard sector is a color singlet. 

\begin{figure}[htb]
    \centering
    \includegraphics[width=0.4\textwidth]{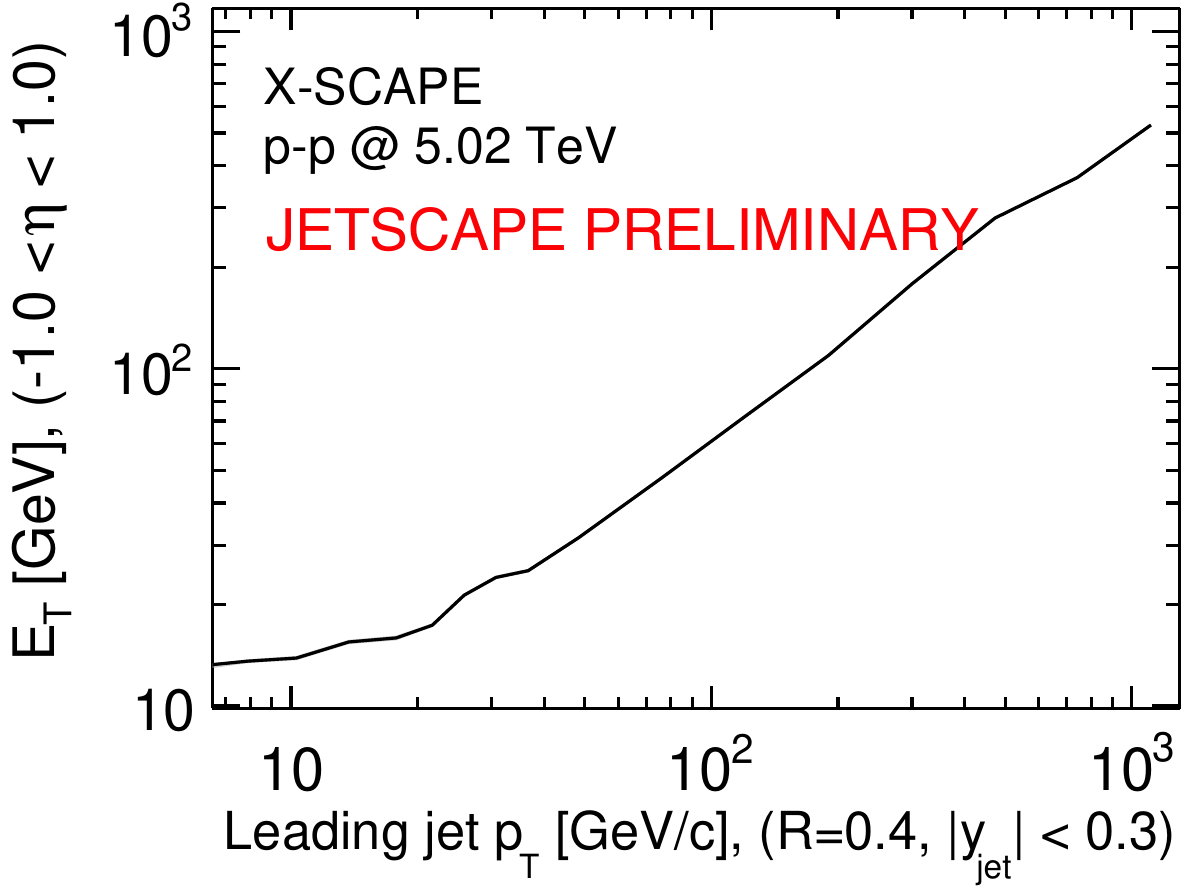}
    \hspace{1cm}
    \includegraphics[width=0.4\textwidth]{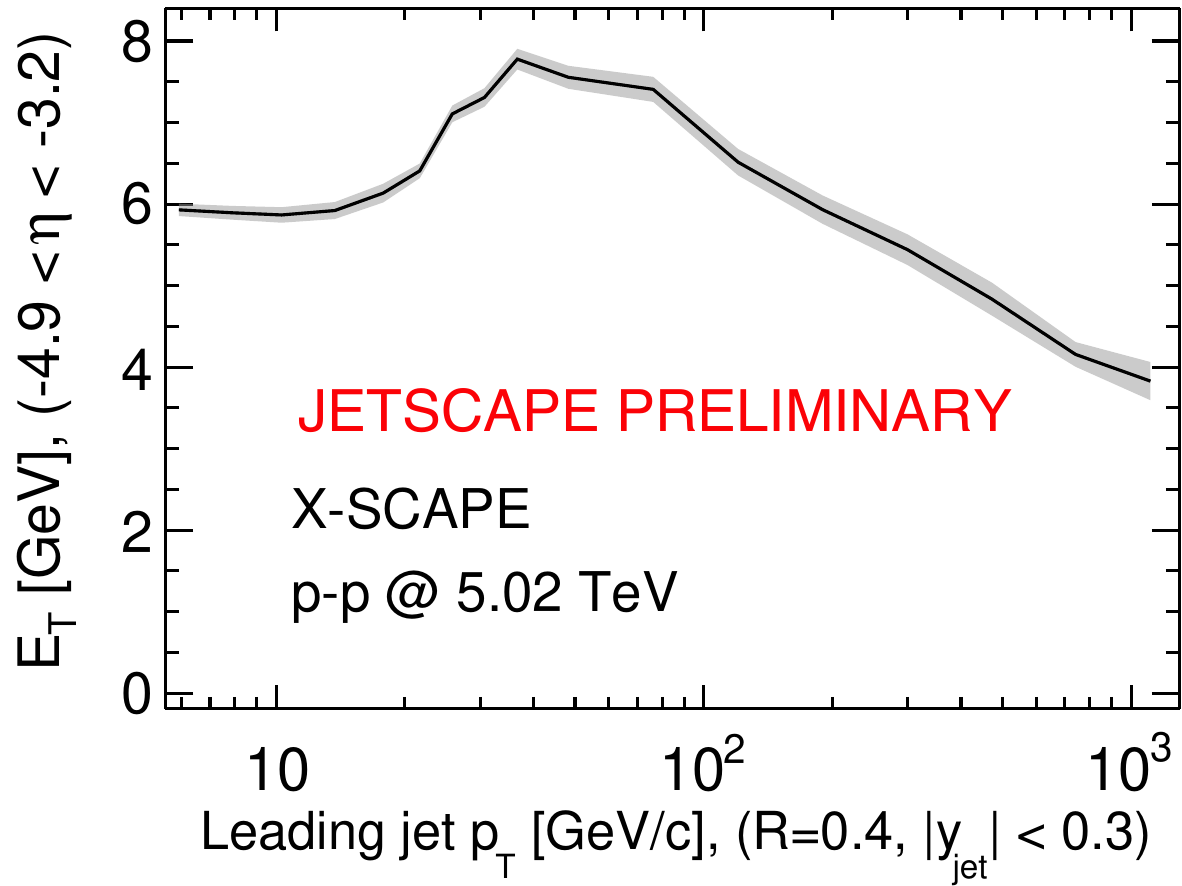}
    \vspace{-0.3cm}
    \caption{The calculated combined (soft + hard process) transverse energy produced in $p$-$p$ collisions at $\sqrt{s}=5.02$~TeV, with $|\eta| < 1$ (left), and $-4.9 < \eta < -3.2$ (right), as a function of the leading jet $p_T$.  }
    \label{fig:pp-multiplicity}
\end{figure}

Both partons and remnants are then conveyed to a second instance of 3D MC-Glauber where their energy and momentum are subtracted from one of the hot spots. The depleted nucleons are then allowed to undergo collision leading to the formation of a bulk medium that lacks the energies of the hard partons constituting the initial state of the hard scattering. The final hadronization has two sources, the hadronization of the bulk medium using Cooper-Frye and the hadronization of the hard sector using the Lund hadronization model. The effect of the energy-momentum depletion from the soft sector can be seen in Fig.~\ref{fig:hydro-output}, where only the hadrons from the bulk sector, over the entire rapidity range, are plotted for different energies within the hard sector, controlled by $\hat{p}_T$, the hardest transverse momentum exchange within the hard sector.

\begin{figure}[htb]
    \centering
    \includegraphics[width=0.4\textwidth]{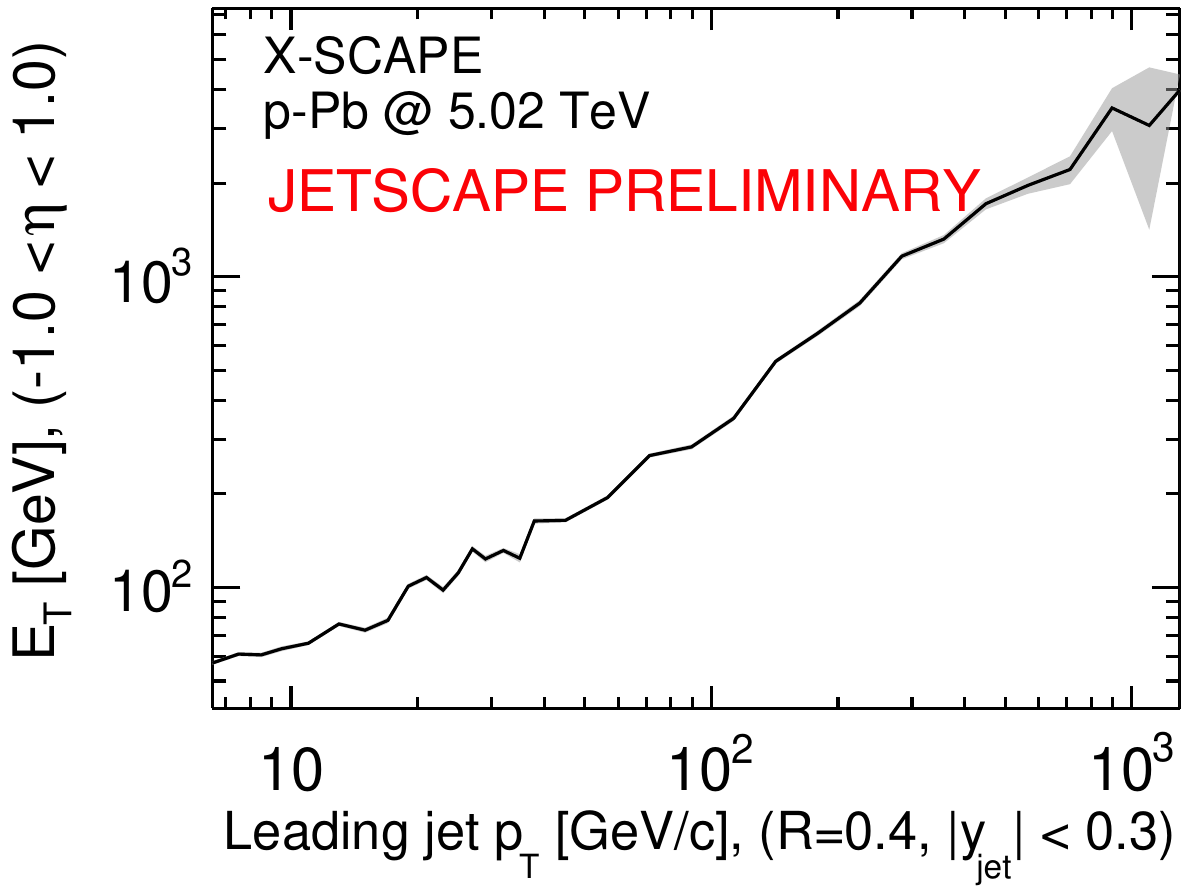}
    \hspace{1cm}
    \includegraphics[width=0.4\textwidth]{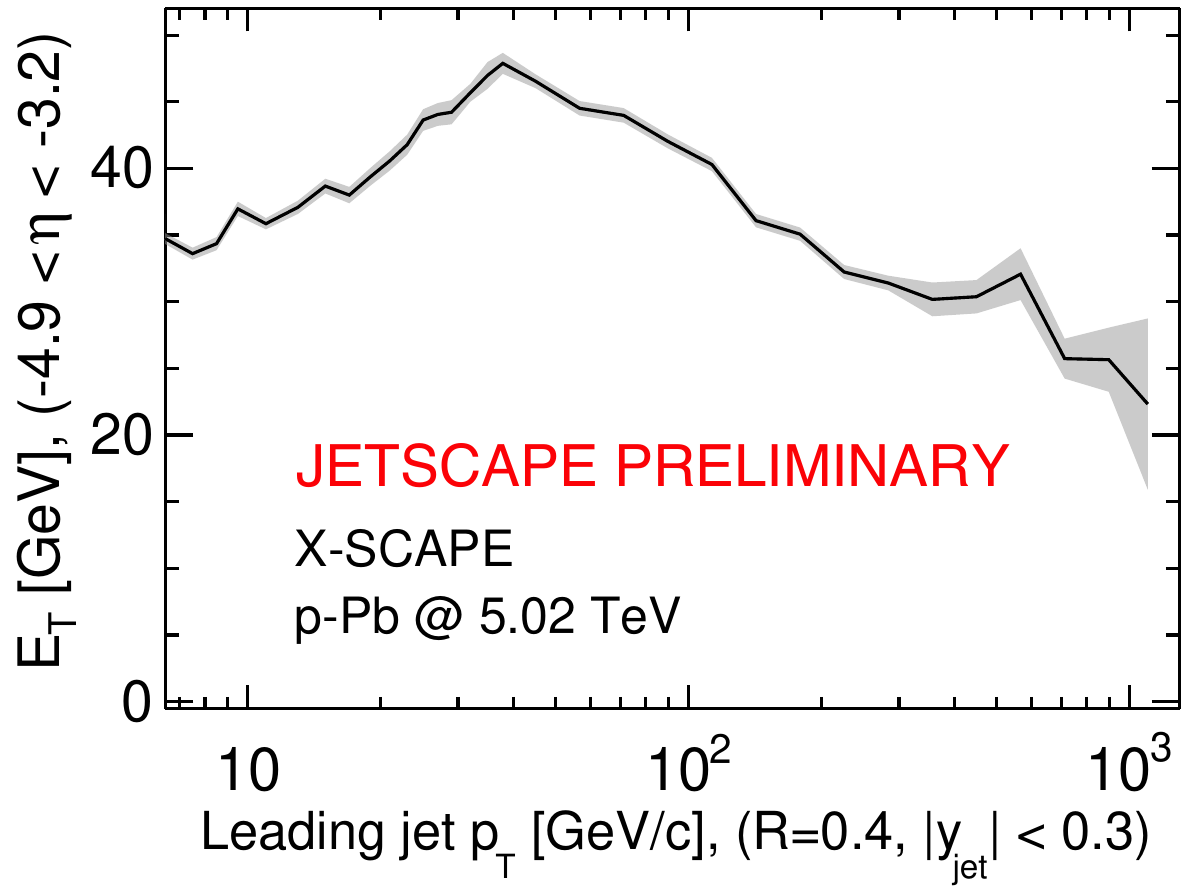}
    \vspace{-0.3cm}
    \caption{Same as Fig.~\ref{fig:pp-multiplicity} above, but for $p$-$Pb$ collisions.}
    \label{fig:p-Pb-multiplicity}
\end{figure}

The value of $\hat{p}_T$ varies over a range from a minimum $\hat{p}_T^0$ to the maximum allowed by kinematics. The value of $\hat{p}_T^0$ controls the energy in the hard sector and thus the energy depletion of the soft sector. We vary this parameter to obtain the best fit with the minimum bias charged pion spectra in $p$-$p$ collisions. This is presented in the left panel of Fig.~\ref{fig:hadron-jet-spectra}. The multiple binary collision enhanced spectra for $p$-$Pb$ is plotted in the same figure. Jets are obtained by clustering hadrons 
from the hard sector and are presented in the right plot of Fig.~\ref{fig:hadron-jet-spectra}.
Nuclear modification factors for both these spectra are presented in Fig.~\ref{fig:hadron-jet-raa}. This is the first time a consistent description of the $R_{AA}$ that explains hadron data from $0 \lesssim p_T \lesssim 20$~GeV, for $p$-$Pb$ collisions has been obtained from one simulation.

With a successful description of the hadron and jet spectrum in $p$-$p$ and $p$-$Pb$ collsions we ask how much does jet production affect multiplicity or transverse energy $E_T$ production at both mid and forward rapidity. This is plotted in Fig.~\ref{fig:pp-multiplicity} for $p$-$p$, and in Fig.~\ref{fig:p-Pb-multiplicity} for $p$-$Pb$. Both these plots show qualitatively similar results. While the particle multiplicity at mid-rapidity increases steadily with the energy of the hardest jet. The multiplicity first rises and then drops at very forward rapidity where the effects of energy conservation between hard and soft sectors are clearly observed.

 \section{Summary}
 \label{sec:summary}
Following these successful preliminary results, the upcoming simulations will feature a Bayesian calibration of the parameters determining the extent of hard and soft sectors. Nuclear modification of the incoming parton distribution functions as well as any possible energy loss of the hard partons within the droplets of the QGP are currently being explored. Also, the effect of initial state fluctuations leading to azimuthal anisotropies in both hard and soft sectors is being studied. 

\noindent
\emph{ACK}: A.M. is supported in part by the NSF (OAC-2004571) and by the DOE (DE-SC0013460). 

\printbibliography[heading=bibintoc,title={References}]

\end{document}